# Layer degree of freedom for excitons in transition metal dichalcogenides

Sarthak Das[∥], Garima Gupta[∥], and Kausik Majumdar[*]

Department of Electrical Communication Engineering, Indian Institute of Science, Bangalore 560012, India

[∥]These authors contributed equally

[*]*Corresponding author, email: kausikm@iisc.ac.in*

**ABSTRACT:** Layered transition metal dichalcogenides (TMDCs) host a variety of strongly bound exciton complexes that control the optical properties in these materials. Apart from spin and valley, layer index provides an additional degree of freedom in a few-layer thick film. Here we show that in a few-layer TMDC film, the wavefunctions of the conduction and valence band edge states contributing to the $K$ ($K'$) valley are spatially confined in the alternate layers - giving rise to direct (quasi-)intra-layer bright exciton and lower-energy inter-layer dark excitons. Depending on the spin and valley configuration, the bright exciton state is further found to be a coherent superposition of two layer-induced states, one (*E-type*) distributed in the even layers and the other (*O-type*) in the odd layers. The intra-layer nature of the bright exciton manifests as a relatively weak dependence of the exciton binding energy on the thickness of the few-layer film, and the binding energy is maintained up to 50 meV in the bulk limit – which is an order of magnitude higher than conventional semiconductors. Fast stokes energy transfer from the intra-layer bright state to the inter-layer dark states provides a clear signature in the layer-dependent broadening of the photoluminescence peak, and plays a key role in the suppression of the photoluminescence intensity observed in TMDCs with thickness beyond monolayer.

## 1. Introduction:

The binding energy of exciton is a strong function of quantum confinement of the electrons and holes. A two-dimensional exciton is thus expected to exhibit stronger binding energy than its three-dimensional counterpart[1]. This, coupled with heavy carrier effective mass[2,3], and small dielectric constant[3–5], results in a large binding energy of excitons in monolayer Transition Metal Dichalcogenide (TMDC) materials[4–9]. This has led to recent efforts in exploring the physics of various exciton complexes including excitons[10], biexcitons[11], trions and their dark states[10,12,13], using monolayer TMDC as a testbed. The inversion symmetry of the crystal is broken in the monolayer limit, and more generally, in TMDCs with odd number of layers, giving rise to rich spin and valley physics[7,14,15]. While exciton complexes have been extensively studied in monolayer TMDCs, the effort in few-layer thick films remains limited[16–23]. This is primarily due to the transition from direct bandgap in monolayer to indirect bandgap in few-layer suggesting fast relaxation of valley carriers from the $K$ ($K'$) points. Also, inversion symmetry is either explicitly restored (in even number of layers) or smears out (in odd number of layers) in multi-layer films, suppressing valley controllability.

On the other hand, few-layer films allow the provision to use layer as an additional degree of freedom. In the 2-H structure of TMDCs, the consecutive layers are rotated by 180° with respect to each other[24]. Consequently, the electrons at the $K$ point in a bilayer system are not allowed to spill over the other layer due to the symmetry of the constituent $d_{z^2}$ orbital contributing to the conduction band. On the other hand, for the holes, there exists a finite inter-layer coupling. However, there is also a large spin splitting in the valence band, the magnitude of which is larger than the inter-layer coupling term. This results in a confinement of the holes to a single layer as well. The spilling of the hole wavefunction to the consecutive layers is particularly weak in W based TMDCs[25,26] compared with Mo based TMDCs owing to larger spin-orbital interaction. Such suppression of inter-layer hopping for both electrons and holes in bilayer TMDCs gives rise to layer pseudospin[24].

However, this argument of single layer confinement is strictly true only at the $K$ ($K'$) points of the Brillouin zone, particularly for few-layer thick film with number of layers more than two. On the contrary, the momentum space distribution of exciton, as predicted from Bethe-Salpeter (BS) equation[27], spreads well beyond the $K$ ($K'$) points, and the wavefunctions spill over to the other layers due to band mixing. In this work, taking the finite momentum space distribution of excitons into account, we generalize the concept of layer degree of freedom for an arbitrary number of layer thickness of WSe$_2$ in the context of the direct exciton to reveal three important properties. First, for a given spin and valley, the layer degree of freedom introduces an additional selection rule for optical brightness. This results from intra- and inter-layer spatial distribution of excitons arising due to electron and hole wavefunctions being distributed either in the odd or in the even layers. Second, the non-radiative scattering from the bright intra-layer to the dark inter-layer states has a clear signature in the layer dependent luminescence linewidth, and plays a key role in luminescence suppression in few-layer TMDC. Third, owing to a pseudo-confinement arising from the quasi-intra-layer nature of the bright exciton, its binding energy is a relatively weak function of thickness of the film, and remains significantly large (~50 meV) even in the bulk limit[28–30] compared to conventional semiconductors[31,32].

**2. Exciton states in few-layer TMDC and their radiative decay:**

To understand the excitonic structure in a few-layer TMDC, we model the exciton using a combination of $\boldsymbol{k}.\boldsymbol{p}$ Hamiltonian and Bethe-Salpeter theory[27,33]. Each layer of WSe$_2$ belongs to the $C_{3h}$ point group at the high symmetry $K$ and $K'$ points of the Brillouin zone, and the W atoms has a trigonal prismatic coordination with the Se atoms. Close to the band edges around the $K$ and $K'$ points in the Brillouin zone (Figure 1a), the bands are contributed primarily from the W $d$ orbitals. The symmetry driven basis states for the conduction band and the valence band edges for the $l^{th}$ layer can be written as[26] $|c\rangle = |5d_{z^2}^l\rangle$ and $|v\rangle = \frac{1}{\sqrt{2}}\left(|5d_{x^2-y^2}^l\rangle + i\tau_z|5d_{xy}^l\rangle\right)$, respectively. Here $\tau_z = \pm 1$ are the $K$ and $K'$ valley indices. For $AB$ stacked TMDC, the adjacent layers are rotated by 180° around the $c$ axis. The Hamiltonian for an $n$-layer WSe$_2$ film is obtained by expanding the monolayer $\mathbf{k}.\mathbf{p}$ Hamiltonian upon incorporating the interlayer coupling of the VBs with the immediate neighbour layers[24]. In **Supplemental Material S1**[34]**,** we show the generalized multi-layer Hamiltonian used in this work. In the same **Supplemental Material**[34], we also show the Hamiltonians for the 2L, 3L and 4L systems explicitly.

An exciton state $|\Psi_s(\mathbf{Q})\rangle$ in an exciton band $s$ at a centre of mass momentum $\mathbf{Q} = \mathbf{k}_e + \mathbf{k}_h$ is a coherent superposition of hole (with crystal momentum $\mathbf{k}_h$) and electron (with crystal momentum $\mathbf{k}_e$) states from band-pairs $(v,c)$ in an $n$-layer system in the reciprocal space and can be written as

$$|\Psi_s(\mathbf{Q})\rangle = \sum_{v,c,\mathbf{k}} \lambda_{v,c,\mathbf{Q}}^{(s)}(\mathbf{k}) |v,\mathbf{k}\rangle|c,\mathbf{k}+\mathbf{Q}\rangle \qquad \ldots (1)$$

$\lambda_{v,c,\mathbf{Q}}^{(s)}(\mathbf{k})$ and the exciton eigen energies $E_{ex}^{(s)}(\mathbf{Q})$ are obtained from the solution of the BS equation[27]:

$$\langle v,c,\mathbf{k},\mathbf{Q}|H|v',c',\mathbf{k}',\mathbf{Q}\rangle = \delta_{vv'}\delta_{cc'}\delta_{\mathbf{kk}'}\left(\varepsilon_{(\mathbf{k}+\mathbf{Q})c} - \varepsilon_{\mathbf{k}v}\right) - (\xi - \varrho)_{vv'}^{cc'}(\mathbf{k},\mathbf{k}',\mathbf{Q}) \quad \ldots (2)$$

Here $\varepsilon$ is the quasiparticle energy eigenvalue obtained by diagonalizing the quasiparticle Hamiltonian in **Supplemental Material S1**[34]. $\varrho$ is the exchange term and we neglect this term since in this work we are primarily interested in the exciton band structure for the direct exciton at $K$ ($K'$), with $\mathbf{Q} \approx \mathbf{0}$. The direct term $\xi$ is evaluated using Keldysh form of coulomb interaction potential[27]: $V_q = \frac{2\pi e^2}{\kappa q(1+r_0 q)}$. The effective dielectric constant $\kappa$ and the characteristic screening length $r_0$ are used as fitting parameters which we vary with the number of layers in the TMDC film.

In the rest of the paper, we only consider the spin allowed, bright transitions, and ignore the selection rule governed dark excitons. Also, we shall limit our discussions to A series excitons only, keeping in mind there exists higher energy exciton series (for example, B series and above). Figure 1a schematically illustrates the one-particle bandstructure for bilayer WSe$_2$. The low-energy bands (1s and 2s) of the $A$ series exciton in the $\mathbf{Q}$ space are shown in Figure 1b. In bilayer, for each of $A_{1s}$ and $A_{2s}$ excitons, there are two layer-induced exciton bands [for example, $A_{1s}^{(1)}$ and $A_{1s}^{(2)}$ for 1s states], which are closely spaced in energy.

Figure 1c shows the light cone, within which the energy and momentum conservation laws are obeyed during an exciton recombination to emit a photon. Thus, any spin allowed bright exciton state with $Q < Q_0$ can emit light by radiative recombination. Owing to the small momentum of the photon compared to the in-plane momentum of the exciton, the light cone constitutes of a small part ($< 0.1\%$) of the Brillouin zone. To compare the strength of the photoluminescence from the different exciton states, we evaluate the radiative decay rate using the following relation[35,36]:

$$\Gamma_R(\mathbf{Q}) = \eta_o \frac{\hbar e^2}{2m_o^2} |\chi_{ex}(\mathbf{Q})|^2 \int_0^\infty dq_z \frac{1}{\sqrt{Q^2+q_z^2}} \times \left(1 + \frac{q_z^2}{Q^2+q_z^2}\right)$$

$$\times \frac{\Gamma(\mathbf{Q})/\pi}{[E_{ex}(\mathbf{Q}) - \hbar c\sqrt{Q^2+q_z^2}] + \Gamma(\mathbf{Q})^2} \quad \ldots (3)$$

Here $\Gamma(\mathbf{Q})$ is the total radiative and non-radiative broadening: $\Gamma(\mathbf{Q}) = \Gamma_R(\mathbf{Q}) + \Gamma_{NR}$. We assume $\Gamma_{NR}$ to be $\mathbf{Q}$-independent for simplicity. See **Supplemental Material S2**[34] for detailed calculation of $\chi_{ex}(\mathbf{Q})$. Note that equation (3) is a self-consistent equation and provides the fundamental radiative broadening of the exciton states when $\Gamma_{NR} = 0$. Figure 1d shows the calculated intrinsic radiative decay rate (for $\Gamma_{NR} = 0$) for both $A_{1s}^{(1)}$ (in black) and $A_{1s}^{(2)}$ (in red) as a function of Q for bilayer WSe$_2$. The implications of the large difference between the two rates which will be discussed later.

### 3. Layer distribution of exciton states – layer-induced bright and dark states:

Fig. 2a schematically shows the conduction and the valence bands of 2L WSe$_2$ for a given spin ($s_z$) and valley ($\tau_z$) index. The top panel describes the doubly degenerate case $\tau_z \cdot s_z = 1$, which includes ($\tau_z = 1$; $s_z = 1$) and ($\tau_z = -1$; $s_z = -1$), while the bottom panel describes the other doubly degenerate case $\tau_z \cdot s_z = -1$, including ($\tau_z = 1$; $s_z = -1$) and ($\tau_z = -1$; $s_z = 1$). At $\mathbf{k} = \mathbf{K}$, owing to the weak inter-layer coupling, states from both conduction and valence bands are confined to either layer L$_1$ or L$_2$, respectively. However, this is not strictly true for $\mathbf{k} \neq \mathbf{K}$. As an example, Fig. 2b shows the layer distribution of the band edge electron and hole states in 2L WSe$_2$ for $\mathbf{k} = \mathbf{K} + \Delta\mathbf{k}$, with $\Delta\mathbf{k} = \frac{2\pi}{a}(0.0033, 0.0033)$ where $a = 3.28$ Å.

The momentum and transition resolved probability distribution ($|\lambda_{v,c,\mathbf{Q}=0}^{(s)}(\mathbf{k})|^2$) of the lowest lying [$A_{1s}^{(1)}$ and $A_{1s}^{(2)}$] and the next higher energy [$A_{2s}^{(1)}$ and $A_{2s}^{(2)}$] direct excitons, as obtained from the BS equation, is shown in Figure 2c-d. Note that, each exciton predominantly consists of a single transition between a specific (v, c) band pair, with negligible contribution from the other transitions. The real space layer distribution of the dominant transition for each exciton state is also schematically shown in Figure 2c-d. The unique layer distribution of the electron and hole basis states governs the formation of intra-layer and inter-layer exciton. For example, for $\tau_z \cdot s_z = 1$, the lower energy exciton [$A_{1s}^{(1)}$] almost entirely arises from C$_1$V$_2$ transition, and hence is an inter-layer exciton as inferred from the top panel of Figure 2b. On the other hand, the higher energy exciton [$A_{1s}^{(2)}$] results primarily from $C_2V_2$ transition and hence forms an intra-layer exciton confined in the bottom layer. On the other hand, for $\tau_z.s_z = -1$, the intra-layer exciton is confined in the top layer, as explained in the bottom panel of Figure 2b. Also note that with an increase in quantum number (from 1s to 2s), the more confined areal distribution of the exciton in the $k$-space suggests a larger spread in real space distribution.

The intra- and inter-layer spatial distributions of the different exciton states are expected to strongly affect their radiative decay. As mentioned earlier, since all these excitons are spin allowed bright states, any exciton with **Q** lying within the light cone (in Figure 1c) can, in principle, recombine radiatively emitting photons in a spontaneous fashion. However, in Figure 1d, we observe that $A_{1s}^{(1)}$ exciton is an order of magnitude weaker compared to the $A_{1s}^{(2)}$ state in terms of light emission due to its inter-layer nature. The primary contributing orbitals (W $5d$) for the excitons exhibit a spatial extent along the out-of-plane ($z$) direction that is much smaller than the interlayer separation (see Figure 2b), suppressing the matrix element for the decay rate in the case of inter-layer exciton. Henceforth, we call these inter-layer states as layer-induced dark excitons. We can thus conclude that the light emission from the 1s state predominantly happens due to the radiative recombination of the intra-layer $A_{1s}^{(2)}$ exciton.

The analysis can be readily extended to the tri-layer (3L) system, and the results are summarized in Figure 3, where three different layer-induced 1s excitons (from the $A$-series) are formed, namely $A_{1s}^{(1)}$, $A_{1s}^{(2)}$ and $A_{1s}^{(3)}$. Figure 3a schematically shows the electronic bandstructure around $K$ and $K'$ points. Similar to bilayer, the components of the eigenstates are significant only in the alternate layers, that is, they are either confined to the even layers or to the odd layers, as illustrated in Figure 3b. The transition and momentum resolved probability distribution of the resulting $A_{1s}$ excitons are explicitly shown in Fig. 3c-e. The probability distributions indicate the dominance of one transition out of 9 possible transitions for an exciton. Similar to the bilayer case, the resulting excitons also follow inter-layer pattern for low energy [$A_{1s}^{(1)}$ and $A_{1s}^{(2)}$] excitons, while intra-layer pattern for the higher energy [$A_{1s}^{(3)}$] state. Interestingly, for $\tau_z \cdot s_z = 1$, the $A_{1s}^{(3)}$ exciton is confined in the middle layer (L$_2$), as shown in the top panel of Figure 3e. However, for the other spin-valley configuration ($\tau_z \cdot s_z = -1$), the $A_{1s}^{(3)}$ exciton is confined to the L$_1$ and L$_3$ (bottom panel of Figure 3e). Thus, it maintains its intra-layer structure, but gets distributed in the odd numbered layers. We term the latter case as quasi-intra-layer exciton. The calculated decay rates of the different exciton states for tri-layer WSe$_2$ are shown in **Supplemental Material S3**[34]. Both types of $A_{1s}^{(3)}$ excitons exhibit more than an order of magnitude higher decay rate compared to the rest, and responsible for photoluminescence.

In Figure 4, we schematically depict the real space layer-resolved distribution of only the bright excitons ($A_{1s}^{(n)}$), for bilayer (2L) to six-layer (6L) thick WSe$_2$ films. We can generalize that for an $n$-layer thick TMDC, there are 2 doubly-degenerate bright (quasi-)intra-layer excitons. The rest $2n-2$ exciton states are inter-layer and hence layer-induced-dark in nature, which are otherwise bright from a conventional selection rule (spin and azimuthal quantum number selection) perspective. Between the two doubly-degenerate bright excitons, one exciton is distributed in the even layers and the other in the odd layers, and we call them as *E-type* (with layer index $l_z = +1$) and *O-type* ($l_z = -1$) exciton, respectively. In the case of 1L and 2L systems, the bright excitons are confined to a single layer. For 3L system, the *E-type* exciton is confined to a single (middle) layer, while the *O-type* one is quasi-intra-layer in nature, being distributed between the top and the bottom layers. For 4L and thicker samples, we only have quasi-intra-layer doubly-degenerate *E-* *type* and *O-type* bright excitons. A careful observation reveals that the spin, valley and layer indices of a bright exciton are coupled by the simple rule $l_z s_z \tau_z = +1$, which dictates the possible quantum states allowed in a few-layer TMDC system.

**4. Experimental evidence and implications:**

We next explore indirect experimental evidences and subsequent implications of the above-mentioned layer distribution of the exciton states. In order to do so, we employ temperature dependent photoluminescence measurement from WSe$_2$ films of varying layer thickness.

**A. Experiment:**

We mechanically exfoliate WSe$_2$ flakes on a clean Si substrate covered with 285 nm thick SiO$_2$. The thickness of the flake is identified by a combination of Raman and AFM. Photoluminescence (PL) measurement is carried out by varying the sample temperature from 3.3 K to room temperature. The pressure of the sample chamber is kept below $10^{-4}$ torr at all measurement temperatures. The PL is collected through a 50X objective with a numerical aperture of 0.5 in confocal mode. The optical power density on the sample is kept below 100 µW to avoid any laser induced heating effect.

Figure 5 summarizes the temperature and thickness dependence of the acquired photoluminescence spectra from WSe$_2$ samples using a 532 nm laser excitation. In Figure 5a, both the neutral ($A_{1s}$) and charged ($A_{1s}^T$) A-series exciton peaks are distinctly visible in the temperature range up to ~90 K for monolayer sample. The red shift of the peak positions with an increase in temperature is due to a corresponding decrease in the quasiparticle bandgap. The weak, but distinct 2s ($A_{2s}$) and 3s ($A_{3s}$) peaks of the A exciton are observed in the zoomed-in Figure 5b around 1.87 eV and 1.93 eV at T = 3.3 K, which smear out as the sample temperature is increased. To confirm that the higher energy peaks originate from the higher order free exciton bright states, we perform polarization resolved photoluminescence measurement at T = 3.3 K. The sample is excited with a $\sigma^+$ circularly polarized light from a 633 nm laser, and the emitted light is passed through a $\sigma^+$ or $\sigma^-$ analyzer. The results for the 1L flake are summarized in Figure 5c. We observe that the $A_{1s}$ exciton peak and the $A_{1s}^T$ trion peak show a degree of circular polarization ($\rho$) of ~8.5% and ~10.2% respectively, where $\rho = \frac{I_{\sigma^+} - I_{\sigma^-}}{I_{\sigma^+} + I_{\sigma^-}}$. In the inset of Figure 5c, we show a magnified portion of the next higher order peak, which shows a strong polarization contrast of ~26%, confirming its $A_{2s}$ assignment. The enhancement of $\rho$ from 1s to 2s is because the 633 nm laser excites the 2s excitons in a near-resonant manner, suppressing the depolarization due to intervalley scattering.

In Figure 5d, we show the acquired PL spectra of WSe$_2$ flakes with varying thickness, namely 1L, 2L, 3L, and 6L, all taken at T = 3.3 K. In the left panel, apart from the neutral and charged exciton peaks, we also observe several peaks at energy lower than trion emission energy. The origin of these lower energy peaks has been previously attributed to defect bound localized excitons[10–12] and multi-particle excitonic states[10–13]. On the other hand, the higher energy peaks, as shown in a magnified energy range in the right panel in Figure 5d, are only distinctly visible for 1L, 2L and 3L cases.

**B. Weak dependence of exciton binding energy on thickness:**

The positions of the $A_{1s}$, $A_{2s}$ and $A_{3s}$ peaks remain almost unaltered (within ~5 meV error bar due to the variation in the individual spectrum obtained from these samples) irrespective of the thickness of the sample. Such layer independence of the $A_{1s}$ exciton peak position has been widely reported previously[37–40].

The (quasi-) intra-layer nature of the bright exciton irrespective of the number of layers in the film forces a spatial pseudo-confinement of the exciton to individual layers. This allows the excitons to retain their two-dimensional character even in multi-layer samples. This effect manifests itself as a weak dependence of the bright exciton binding energy on the number of layers of the film. The bright excitons being accessible by photoluminescence experiment, allows us to readily verify this hypothesis. The exciton emission energies, calculated from Equation 2 for different layers, are plotted as a function of the quantum numbers in Figure 6a as the open symbols, which are in good agreement with the experimental data, shown by the solid symbols. The insets show zoomed-in views of the data from individual layers. The corresponding continuum levels obtained from the BS equation for different layer numbers are also shown in the same figure by solid horizontal lines. The corresponding binding energy of the different exciton states is then calculated by taking the difference between the continuum level and the emission energy (obtained from both photoluminescence spectra as well as BS equation) and plotted as a function of the thickness of the $WSe_2$ flake in Figure 6b. The agreement between BS theory and experiment is quite remarkable. The observation of the weak dependence of the exciton binding energy on $WSe_2$ film thickness is in stark contrast with a conventional semiconductor when the out of plane quantum confinement is relaxed. Also, the binding energy of the exciton for bulk TMDC is about 50 meV, which was measured long back[28–30]. This is about an order of magnitude higher than typical exciton binding energies of III-V semiconductor samples[31,32]. The retention of the large binding energy in the bulk limit is another implication of such quasi-intra-layer configuration of the bright excitons in TMDCs, which maintains a quasi-two-dimensional nature due to layer confinement even in thick samples. A summary of the layer dependence on the energy and the binding energy of different excitonic states is provided in **Supplemental Material S4**[34].

## C. Layer dependent exciton linewidth broadening:

Using a Voigt fit to the exciton peaks for samples with varying layer thickness, we deconvolute the homogeneous (Lorentzian) and the inhomogeneous (Gaussian, shown in **Supplemental Materials S5**[34]) components of the exciton photoluminescence linewidth. The extracted homogeneous linewidth is found to increase monotonically as a function of number of layers ($n$) in Figure 7a (green symbols). The total homogeneous linewidth is a result of both radiative and non-radiative dephasing processes. Using the self-consistency of Equation 3, we deconvolute the corresponding non-radiative part $[2\Gamma_{NR}(n)]$ of the homogeneous broadening as a function of $n$ from the photoluminescence homogeneous linewidth[36]. $2\Gamma_{NR}(n)$ is found to increase linearly with $n$. As the excitation density was maintained low ($< 10^9$ cm$^{-2}$) during measurements, the exciton-exciton scattering induced dephasing[41] is small, and the exciton-phonon scattering is the dominating non-radiative dephasing process in a monolayer sample in our experiment. For $n \geq 2$, apart from the exciton-phonon scattering within the bright $A_{1s}^{(n)}$ band, scattering to the indirect valleys ($\boldsymbol{\Gamma}$ and $\boldsymbol{\Lambda}$) and to the lower energy inter-layer dark states are the additional non-radiative dephasing mechanisms. We assume that the phonon scattering within the $A_{1s}^{(n)}$ band is independent of layer number, and therefore, is equal to the monolayer non-radiative linewidth ($\Gamma_0$). The layer dependence of the non-radiative linewidth can then be given by

$$\Gamma_{NR}(n) = \Gamma_0 + \Gamma_I + \sum_{i=1}^{n-1} \Gamma_{A_{1s}^{(n)} \to A_{1s}^{(i)}} \quad ; n > 1 \quad \ldots (4)$$

Here, $\Gamma_I$ quantifies the lumped effect of dephasing due to exciton-phonon scattering to the indirect valleys. Due to large inter-valley momentum mismatch, $\Gamma_I$ is expected to be small compared to intra-valley scattering rates. In the last term, $\Gamma_{A_{1s}^{(n)} \to A_{1s}^{(i)}}$ is the scattering of the bright (quasi-)intra-layer $A_{1s}^{(n)}$ exciton to the $i^{th}$ dark inter-layer $A_{1s}^{(i)}$ exciton, maintaining both their spin and valley indices (that is, conserving total angular momentum). Note that $A_{1s}^{(i)}$ ranges from $A_{1s}^{(1)} \rightarrow A_{1s}^{(n-1)}$, and this results in a proportionately increasing number of scattering channels as the number of layers increases (see Figure 7b for 2L and 3L cases). For a first order estimate, we assume the same scattering rate [denoted by $\Gamma_{A_{1s}^{(B)} \rightarrow A_{1s}^{(D)}}$] from bright $A_{1s}^{(n)}$ to any of the $i^{th}$ lower energy inter-layer dark state. We can then rewrite Equation 4 as

$$\Gamma_{NR}(n) = \Gamma_0 + \Gamma_I + (n-1)\Gamma_{A_{1s}^{(B)} \rightarrow A_{1s}^{(D)}} \qquad ; n > 1 \qquad \ldots (5)$$

This explains the linear increment in the non-radiative exciton linewidth as the number of layers increases.

**D. Photoluminescence suppression beyond monolayer:**

$\Gamma_{A_{1s}^{(B)} \rightarrow A_{1s}^{(D)}}$ is extracted from the slope of the linear fit from Figure 7a, and is found to be ~2.25 meV, which translates to a scattering rate of $3.4 \times 10^{12}$ s$^{-1}$ per channel. This is on the order of radiative decay rate of the bright exciton (see Figure 1d) as well as the carrier transfer to indirect valleys[42]. Equation (5) suggests that with an increase in the number of layers, the total non-radiative decay rate due to intra-layer to inter-layer stokes energy transfer increases proportionately. Since the inter-layer states do not contribute to the luminescence, such non-radiative scattering competes with the exciton radiative decay process. This suggests that apart from carrier transfer to the indirect valleys, the fast scattering to the inter-layer dark states also plays a key role in suppressing luminescence in few-layer TMDCs.

**Conclusions:**

In summary, the symmetry driven even- and odd-layer distribution of the band edge states close to the zone corner forces intra- (or quasi-intra-) and inter-layer distribution of excitons in few-layer

TMDCs. The intra-layer exciton states exhibit more than an order of magnitude higher radiative decay rate compared to the inter-layer states, and hence only these excitons contribute to the luminescence. These bright intra-layer excitons can further be classified into *E-type* and *O-type* excitons (denoted as layer index), depending on their spatial layer distribution over either even or odd numbered layers, respectively. This layer index ($l_z$) is coupled to the spin ($s_z$) and valley ($\tau_z$) indices by the rule $l_z s_z \tau_z = +1$. Such unique layer distribution has direct implication in maintaining large exciton binding energy in TMDCs up to the bulk limit. Further, the layer index (*E* or *O*) can be treated as an additional degree of freedom of the exciton quantum state in a few-layer system, and can be used for quantum information manipulation.

**ACKNOWLEDGMENTS**

K. M. acknowledges the support a grant from Indian Space Research Organization (ISRO), grants under Ramanujan Fellowship, Early Career Award, and Nano Mission from the Department of Science and Technology (DST), Government of India, and support from MHRD, MeitY and DST Nano Mission through NNetRA.

**Competing interests**

The authors declare no competing financial or non-financial interests.

**References:**

(1) X. L. Yang, S. H. Guo, F. T. Chan, K. W. W. and W. Y. C. Analytic Solution of a Two-Dimensional Hydrogen Atom. I. Nonrelativistic Theory. *Phys. Rev. A* **1991**, *43* (3), 1186–1196.

(2) Majumdar, K.; Hobbs, C.; Kirsch, P. D. Benchmarking Transition Metal Dichalcogenide MOSFET in the Ultimate Physical Scaling Limit. *IEEE Electron Device Lett.* **2014**, *35* (3), 402–404. https://doi.org/10.1109/LED.2014.2300013.

(3) Ramasubramaniam, A. Large Excitonic Effects in Monolayers of Molybdenum and Tungsten Dichalcogenides. *Phys. Rev. B - Condens. Matter Mater. Phys.* **2012**, *86* (11), 1–6. https://doi.org/10.1103/PhysRevB.86.115409.

(4) He, K.; Kumar, N.; Zhao, L.; Wang, Z.; Mak, K. F.; Zhao, H.; Shan, J. Tightly Bound Excitons in Monolayer WSe2. *Phys. Rev. Lett.* **2014**, *113* (2), 1–5. https://doi.org/10.1103/PhysRevLett.113.026803.

(5) Komsa, H. P.; Krasheninnikov, A. V. Effects of Confinement and Environment on the Electronic Structure and Exciton Binding Energy of MoS2 from First Principles. *Phys. Rev. B - Condens. Matter Mater. Phys.* **2012**, *86* (24), 1–6. https://doi.org/10.1103/PhysRevB.86.241201.

(6) Chernikov, A.; Berkelbach, T. C.; Hill, H. M.; Rigosi, A.; Li, Y.; Aslan, O. B.; Reichman, D. R.; Hybertsen, M. S.; Heinz, T. F. Exciton Binding Energy and Nonhydrogenic Rydberg Series in Monolayer WS2. *Phys. Rev. Lett.* **2014**, *113* (7), 1–5. https://doi.org/10.1103/PhysRevLett.113.076802.

(7) Xu, X.; Yao, W.; Xiao, D.; Heinz, T. F. Spin and Pseudospins in Layered Transition Metal Dichalcogenides. *Nat. Phys.* **2014**, *10* (5), 343–350. https://doi.org/10.1038/nphys2942.

(8) Shi, H.; Pan, H.; Zhang, Y. W.; Yakobson, B. I. Quasiparticle Band Structures and Optical Properties of Strained Monolayer MoS2 and WS2. *Phys. Rev. B - Condens. Matter Mater. Phys.* **2013**, *87* (15), 1–8. https://doi.org/10.1103/PhysRevB.87.155304.

(9) Gupta, G.; Kallatt, S.; Majumdar, K. Direct Observation of Giant Binding Energy Modulation of Exciton Complexes in Monolayer MoSe2. *Phys. Rev. B* **2017**, *96* (8), 081403. https://doi.org/10.1103/PhysRevB.96.081403.

(10) Jones, A. M.; Yu, H.; Ghimire, N. J.; Wu, S.; Aivazian, G.; Ross, J. S.; Zhao, B.; Yan, J.; Mandrus, D. G.; Xiao, D.; et al. Optical Generation of Excitonic Valley Coherence in Monolayer WSe2. *Nat. Nanotechnol.* **2013**, *8* (9), 634–638. https://doi.org/10.1038/nnano.2013.151.

(11) You, Y.; Zhang, X.-X.; Berkelbach, T. C.; Hybertsen, M. S.; Reichman, D. R.; Heinz, T. F. Observation of Biexcitons in Monolayer WSe2. *Nat. Phys.* **2015**, *11* (6), 477–481. https://doi.org/10.1038/nphys3324.

(12) Zhang, X.-X.; Cao, T.; Lu, Z.; Lin, Y.-C.; Zhang, F.; Wang, Y.; Li, Z.; Hone, J. C.; Robinson, J. A.; Smirnov, D.; et al. Magnetic Brightening and Control of Dark Excitons in Monolayer WSe2. *Nat. Nanotechnol.* **2017**, No. June, 1–7. https://doi.org/10.1038/nnano.2017.105.

(13) Wang, G.; Bouet, L.; Lagarde, D.; Vidal, M.; Balocchi, A.; Amand, T.; Marie, X.; Urbaszek, B. Valley Dynamics Probed through Charged and Neutral Exciton Emission in Monolayer WSe2. *Phys. Rev. B - Condens. Matter Mater. Phys.* **2014**, *90* (7), 1–6.

https://doi.org/10.1103/PhysRevB.90.075413.

(14) Yao, W.; Xiao, D.; Niu, Q. Valley-Dependent Optoelectronics from Inversion Symmetry Breaking. *Phys. Rev. B - Condens. Matter Mater. Phys.* **2008**, *77* (23), 1–7. https://doi.org/10.1103/PhysRevB.77.235406.

(15) Srivastava, A.; Sidler, M.; Allain, A. V.; Lembke, D. S.; Kis, A.; Imamoglu, A. Valley Zeeman Effect in Elementary Optical Excitations of a Monolayer WSe2. **2014**, *11* (February), 141–147. https://doi.org/10.1038/nphys3203.

(16) Dai, X.; Li, W.; Wang, T.; Wang, X.; Zhai, C. Bandstructure Modulation of Two-Dimensional WSe2 by Electric Field. *J. Appl. Phys.* **2015**, *117* (8). https://doi.org/10.1063/1.4907315.

(17) Li, D.; Cheng, R.; Zhou, H.; Wang, C.; Yin, A.; Chen, Y.; Weiss, N. O.; Huang, Y.; Duan, X. Electric-Field-Induced Strong Enhancement of Electroluminescence in Multilayer Molybdenum Disulfide. *Nat. Commun.* **2015**, *6* (May), 7509. https://doi.org/10.1038/ncomms8509.

(18) Palummo, M.; Bernardi, M.; Grossman, J. C. Exciton Radiative Lifetimes in Two-Dimensional Transition Metal Dichalcogenides. *Nano Lett.* **2015**, *15* (5), 2794–2800. https://doi.org/10.1021/nl503799t.

(19) Zhang, Y.; Chang, T.-R.; Zhou, B.; Cui, Y.-T.; Yan, H.; Liu, Z.; Schmitt, F.; Lee, J.; Moore, R.; Chen, Y.; et al. Direct Observation of the Transition from Indirect to Direct Bandgap in Atomically Thin Epitaxial MoSe2. **2014**, *9* (February), 111–115. https://doi.org/10.1038/nnano.2013.277.

(20) Brumme, T.; Calandra, M.; Mauri, F. First-Principles Theory of Field-Effect Doping in Transition-Metal Dichalcogenides: Structural Properties, Electronic Structure, Hall Coefficient, and Electrical Conductivity. *Phys. Rev. B - Condens. Matter Mater. Phys.* **2015**, *91* (15), 1–19. https://doi.org/10.1103/PhysRevB.91.155436.

(21) Li, X.; Zhang, F.; Niu, Q. Unconventional Quantum Hall Effect and Tunable Spin Hall Effect in Dirac Materials: Application to an Isolated MoS2 Trilayer. *Phys. Rev. Lett.* **2013**, *110* (6), 1–5. https://doi.org/10.1103/PhysRevLett.110.066803.

(22) Roldán, R.; López-Sancho, M. P.; Guinea, F.; Cappelluti, E.; Silva-Guillén, J. A.; Ordejón, P. Momentum Dependence of Spin–orbit Interaction Effects in Single-Layer and Multi-Layer Transition Metal Dichalcogenides. *2D Mater.* **2014**, *1* (3), 034003. https://doi.org/10.1088/2053-1583/1/3/034003.

(23) Roldán, R.; Silva-Guillén, J. A.; López-Sancho, M. P.; Guinea, F.; Cappelluti, E.; Ordejón, P. Electronic Properties of Single-Layer and Multilayer Transition Metal Dichalcogenides MX2 (M = Mo, W and X = S, Se). *Ann. Phys.* **2014**, *526* (9–10), 347–357. https://doi.org/10.1002/andp.201400128.

(24) Gong, Z.; Liu, G.-B.; Yu, H.; Xiao, D.; Cui, X.; Xu, X.; Yao, W. Magnetoelectric Effects and Valley-Controlled Spin Quantum Gates in Transition Metal Dichalcogenide Bilayers. *Nat. Commun.* **2013**, *4* (May), 1–6. https://doi.org/10.1038/ncomms3053.

(25) Latzke, D. W.; Zhang, W.; Suslu, A.; Chang, T. R.; Lin, H.; Jeng, H. T.; Tongay, S.; Wu, J.; Bansil, A.; Lanzara, A. Electronic Structure, Spin-Orbit Coupling, and Interlayer Interaction in Bulk MoS2 and WS2. *Phys. Rev. B - Condens. Matter Mater. Phys.* **2015**, *91* (23), 1–6. https://doi.org/10.1103/PhysRevB.91.235202.

(26) Xiao, D.; Liu, G. Bin; Feng, W.; Xu, X.; Yao, W. Coupled Spin and Valley Physics in Monolayers of MoS 2 and Other Group-VI Dichalcogenides. *Phys. Rev. Lett.* **2012**, *108* (19), 1–5. https://doi.org/10.1103/PhysRevLett.108.196802.

(27) Wu, F.; Qu, F.; Macdonald, A. H. Exciton Band Structure of Monolayer MoS2. *Phys. Rev. B - Condens. Matter Mater. Phys.* **2015**, *91* (7), 1–8. https://doi.org/10.1103/PhysRevB.91.075310.

(28) Beal, A. R.; Knights, J. C.; Liang, W. Y. Transmission Spectra of Some Transition Metal Dichalcogenides : 11. Group VIA : Trigonal Prismatic Coordination. *J. Phys. J. Phys. C Solid State Phys* **1972**, *5*.

(29) Beal, A. R.; Liang, W. Y. Excitons in 2H-WSe2 and 3R-WS2. *J. Phys. C. Solid State Phys.* **1976**, *9*, 2459–2466.

(30) Wilson, J. A.; Yoffe, A. D. The Transition Metal Dichalcogenides Discussion and Interpretation of the Observed Optical, Electrical and Structural Properties. *Adv. Phys.* **1969**, *18* (73), 193–335. https://doi.org/10.1080/00018736900101307.

(31) Greene, R. L.; Bajaj, K. K.; Phelps, D. E. Energy Levels of Wannier Excitons in GaAs-Ga1-XAlxAs Quantum-Well Structures. *Phys. Rev. B* **1984**, *29* (4), 1807–1812. https://doi.org/10.1103/PhysRevB.29.1807.

(32) Miller, R. C.; Kleinman, D. A.; Tsang, W. T.; Gossard, A. C. Observation of the Excited Level of Excitons in GaAs Quantum Wells. *Phys. Rev. B* **1981**, *24* (2), 1134–1136. https://doi.org/10.1103/PhysRevB.24.1134.

(33) Qiu, D. Y.; Cao, T.; Louie, S. G. Nonanalyticity, Valley Quantum Phases, and Lightlike

Exciton Dispersion in Monolayer Transition Metal Dichalcogenides: Theory and First-Principles Calculations. *Phys. Rev. Lett.* **2015**, *115* (17), 1–5. https://doi.org/10.1103/PhysRevLett.115.176801.

(34) Das, S.; Gupta, G.; Majumdar, K. Supplemental Material for : Layer Degree of Freedom for Excitons in Transition Metal Dichalcogenides.

(35) Wang, H.; Bang, J.; Sun, Y.; Liang, L.; West, D.; Meunier, V.; Zhang, S. The Role of Collective Motion in the Ultrafast Charge Transfer in van Der Waals Heterostructures. *Nat. Commun.* **2016**, *7*, 11504.

(36) Gupta, G.; Majumdar, K. Fundamental Exciton Linewidth Broadening in Monolayer Transition Metal Dichalcogenides. **2019**, *085412* (September 2018), 1–9. https://doi.org/10.1103/PhysRevB.99.085412.

(37) Maciej R. Molas, Karol Nogajewski, Artur O. Slobodeniuk, Johannes Binder, Miroslav Bartos, and M. P. Optical Response of Monolayer, Few-Layer and Bulk Tungsten Disulfide. *Nanoscale* **2017**, *7*, 10421–10429. https://doi.org/10.1039/b000000x/.

(38) Arora, A.; Koperski, M.; Nogajewski, K.; Marcus, J.; Faugeras, C.; Potemski, M. Excitonic Resonances in Thin Films of WSe2: From Monolayer to Bulk Material. *Nanoscale* **2015**, *7* (23), 10421–10429. https://doi.org/10.1039/c5nr01536g.

(39) Zeng, H.; Liu, G. Bin; Dai, J.; Yan, Y.; Zhu, B.; He, R.; Xie, L.; Xu, S.; Chen, X.; Yao, W.; et al. Optical Signature of Symmetry Variations and Spin-Valley Coupling in Atomically Thin Tungsten Dichalcogenides. *Sci. Rep.* **2013**, *3*, 2–6. https://doi.org/10.1038/srep01608.

(40) Zhao, W.; Ghorannevis, Z.; Chu, L.; Toh, M.; Kloc, C.; Tan, P.-H.; Eda, G. Evolution of

Electronic Structure in Atomically Thin Sheets of WS$_2$ and WSe$_2$. *ACS Nano* **2012**, *7* (1), 791–797.

(41) Moody, G.; Kavir Dass, C.; Hao, K.; Chen, C.-H.; Li, L.-J.; Singh, A.; Tran, K.; Clark, G.; Xu, X.; Berghäuser, G.; et al. Intrinsic Homogeneous Linewidth and Broadening Mechanisms of Excitons in Monolayer Transition Metal Dichalcogenides. *Nat. Commun.* **2015**, *6* (May), 8315. https://doi.org/10.1038/ncomms9315.

(42) Bertoni, R.; Nicholson, C. W.; Waldecker, L.; Hübener, H.; Monney, C.; De Giovannini, U.; Puppin, M.; Hoesch, M.; Springate, E.; Chapman, R. T.; et al. Generation and Evolution of Spin-, Valley-, and Layer-Polarized Excited Carriers in Inversion-Symmetric WSe2. *Phys. Rev. Lett.* **2016**, *117* (27), 1–5. https://doi.org/10.1103/PhysRevLett.117.277201.

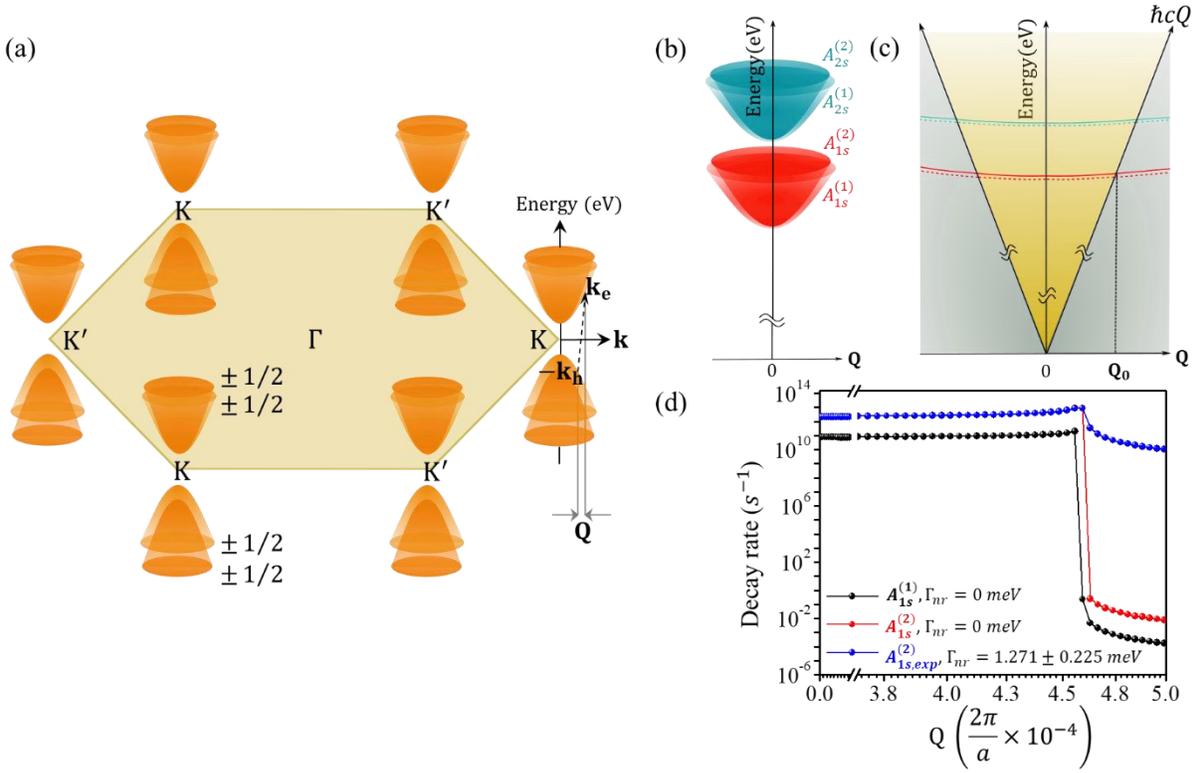

**Figure 1. Electronic and excitonic band structure in bilayer TMDCs.** (a) Band diagram showing the conduction band (CB) and valence band (VB) around the corners of the hexagonal Brillouin zone in bilayer TMDCs. Each band is spin degenerate. (b) Exciton band dispersion of $1s$ and $2s$ excitons with its centre-of-mass momentum ($Q$). Contrary to a monolayer system, there are two possible $1s$ states arising from layer degree of freedom. (c) Light cone for excitons, where $Q_0$ is the edge of the conventional light cone and the boundary of the light cone is given by the light line $\hbar cQ$. The excitons occupying the lower energy $A_{1s}^{(1)}$ and $A_{2s}^{(1)}$ bands (dashed line) are dark in nature because of their inter-layer character, whereas, the higher energy $A_{1s}^{(2)}$ and $A_{2s}^{(2)}$ excitons (solid line) are bright due to their intra-layer character. (d) Radiative decay rate variation of the two $1s$ exciton bands in (b) with $Q$ (in units of $10^{-4}\frac{2\pi}{a}$) on varying the non-radiative linewidth of the exciton band. In the absence of non-radiative scattering ($\Gamma_{NR} = 0$), the decay rate of the $A_{1s}^{(2)}$ exciton (red) is roughly two orders of magnitude larger than the $A_{1s}^{(1)}$ exciton (black), showing that the lower energy $A_{1s}^{(1)}$ exciton is radiatively inefficient compared to the higher energy $A_{1s}^{(2)}$ state. As $\Gamma_{NR}$ increases, the decay rate outside the light cone boundary $Q_0$ for the bright $A_{1s}^{(2)}$ exciton increases due to enhanced participation of the broadened exciton states above the light line.

**Figure 2. Exciton formation in 2L WSe₂.** (a) Schematic of layer induced bands at the zone corner for different spin ($s_z$) and valley ($\tau_z$) configurations. $V_i$ and $C_i$ correspond to the i$^{th}$ valence and conduction band, respectively. The spin and valley configuration for the top and bottom rows has been followed in the subsequent figures also. (b) Real space distribution of different bands at $\boldsymbol{k} = \boldsymbol{K} + \Delta \boldsymbol{k}$, with $\Delta \boldsymbol{k} = \frac{2\pi}{a}(0.0033, 0.0033)$ with the same spin and valley configuration indicated in in figure (a). The conduction and valence bands are shown in the left and right panels respectively. The physical locations of the layers are shown in the middle. (c-d) $\boldsymbol{k}$-space distribution of (c) the $A_{1s}$ exciton and (d) the $A_{2s}$ exciton, for all the possible individual transitions for a bilayer system. The two different $A_{1s}$ states have been indicated according to their dark (in grey boxes) and bright (in yellow boxes) nature. The corresponding real space layer-resolved distribution for each exciton configuration (green and copper spheres indicate the hole and the electron, respectively) is illustrated schematically above the top panel and below the bottom panels. $A_{2s}$ excitons are more separated in real space while more confined in the $k$-space.

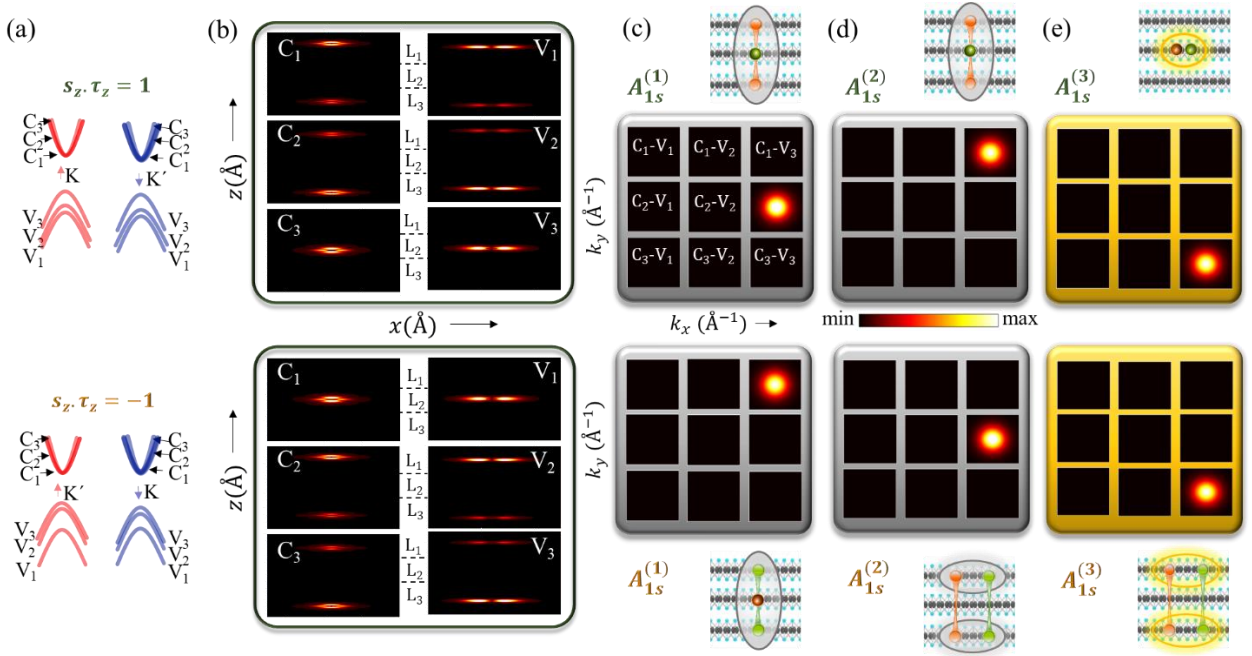

**Figure 3. Exciton formation in 3L WSe$_2$.** (a) Schematic of layer induced bands at the zone corner for different spin ($s_z$) and valley ($\tau_z$) configurations. V$_i$ and C$_i$ correspond to the i$^{th}$ valence and conduction band, respectively. (b) Real space distribution of different bands at $\mathbf{k} = \mathbf{K} + \Delta \mathbf{k}$, with $\Delta \mathbf{k} = \frac{2\pi}{a}(0.0033, 0.0033)$ with the same spin valley configuration indicated in in figure (a). The conduction and valence bands are shown in the left and right panels, respectively. The physical locations of the layers are shown in the middle. (c-e) $k$-space distribution of the $A_{1s}$ exciton for all the possible individual transitions for a tri-layer system. The three different $A_{1s}$ states have been indicated according to their dark (in grey boxes) in (c-d) and bright (in yellow boxes) nature in (e). The corresponding real space layer-resolved distribution for each exciton configuration (green and copper spheres indicate the hole and the electron, respectively) is illustrated schematically above the top panel and below the bottom panels. Depending on layer distribution, low energy $A_{1s}^{(1)}$ and $A_{1s}^{(2)}$ forms the inter-layer excitons in (c-d) while the high energy $A_{1s}^{(3)}$ forms the intra-layer excitons in (e).

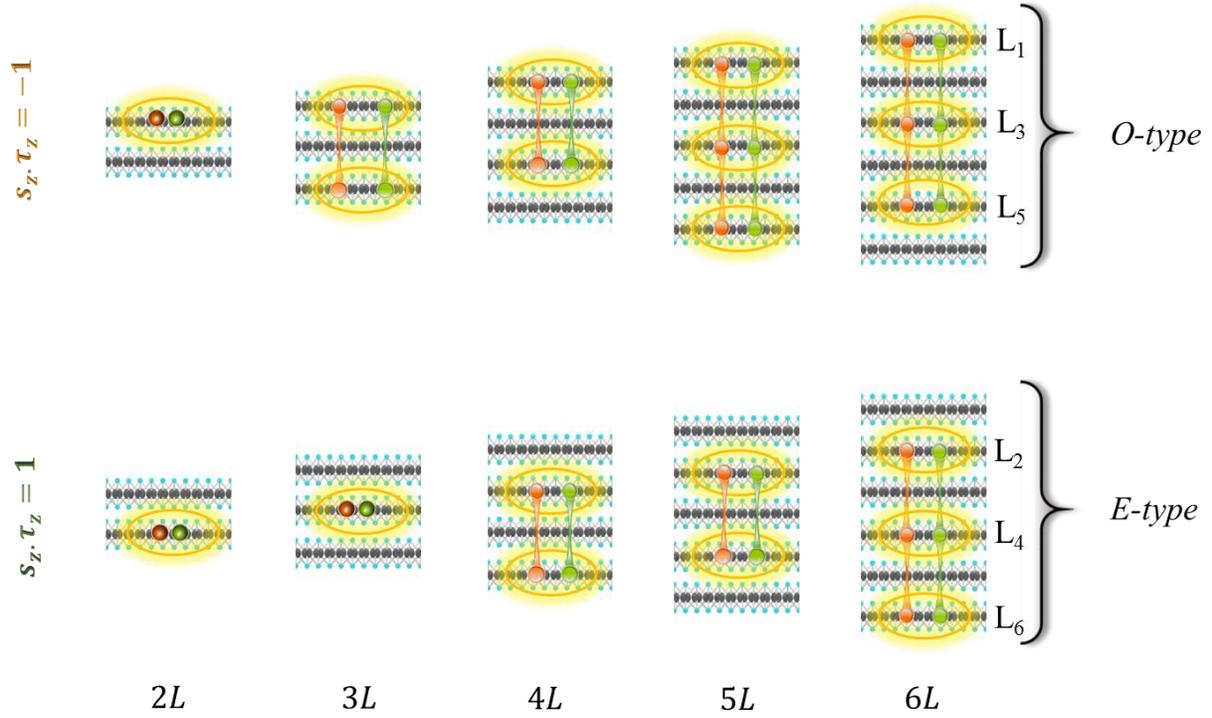

**Figure 4. Layer distribution of bright excitons.** Schematic for the real space layer-resolved distribution of different intra-layer bright exciton states for bilayer (2L) to six-layers (6L) for two doubly degenerate configurations. Depending on their layer distribution, these are classified as *O-type* (distributed in the odd numbered layers, top panel, layer index $l_z = -1$) and *E-type* (distributed in the even numbered layers, bottom panel, layer index $l_z = +1$) excitons.

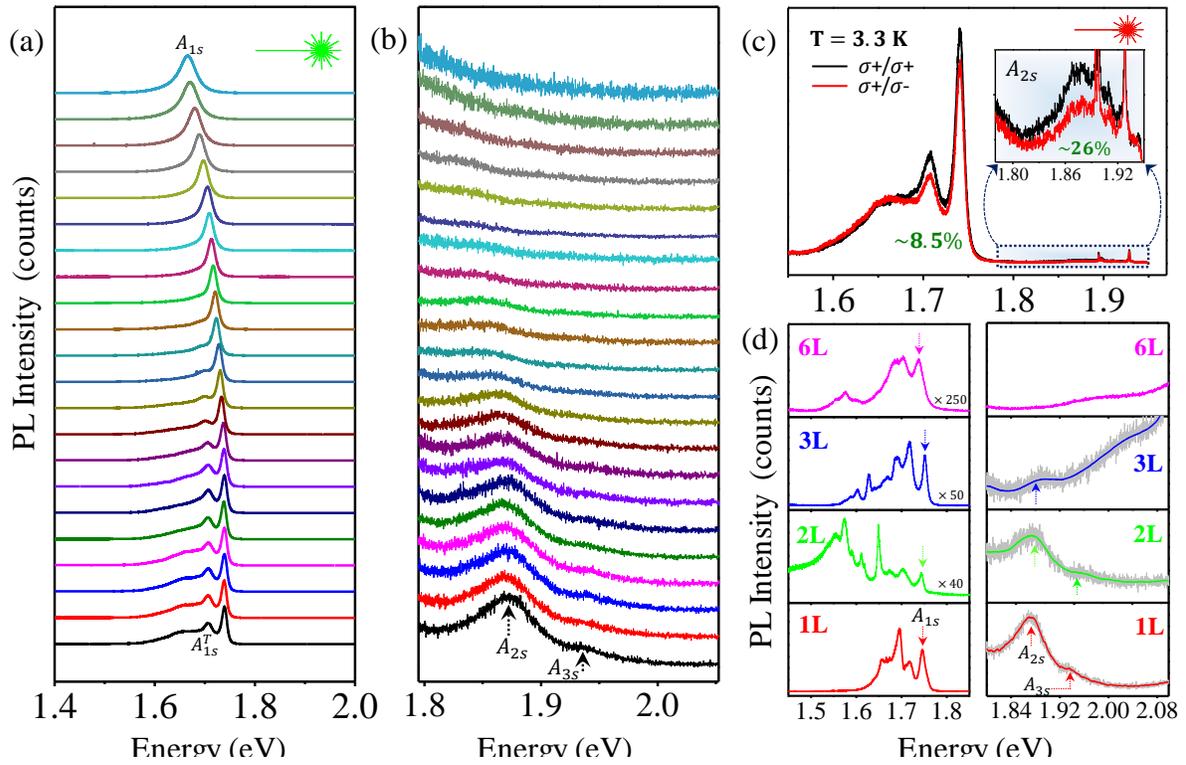

**Figure 5. Exciton states in WSe$_2$ probed through photoluminescence.** (a) PL intensity variation of 1L WSe$_2$ at sample temperature ranging from 3.3 K to 295 K with 532 nm laser excitation. The corresponding exciton ($A_{1s}$) and trion ($A_{1s}^T$) peaks are indicated. (b) The magnified view of the higher order exciton ($A_{2s}$ and $A_{3s}$) peaks with increasing temperature, as indicated by arrows. (c) Circular polarization resolved PL spectra for monolayer WSe$_2$ with a 633 nm laser excitation at 3.3 K with a polarization contrast ($\rho$) of ~8.5% for $A_{1s}$. Inset. The degree of polarization of $A_{2s}$ peak is around 26%. (d) PL spectra of 1L, 2L, 3L, and 6L WSe$_2$, with 532 nm laser excitation at 3.3 K in the left panel. PL spectra showing higher order exciton peaks ($A_{2s}$ and $A_{3s}$) for the same samples in (d) in the right panel. The higher order peaks (indicated by arrows) are discernible only for 1L, 2L and 3L samples.

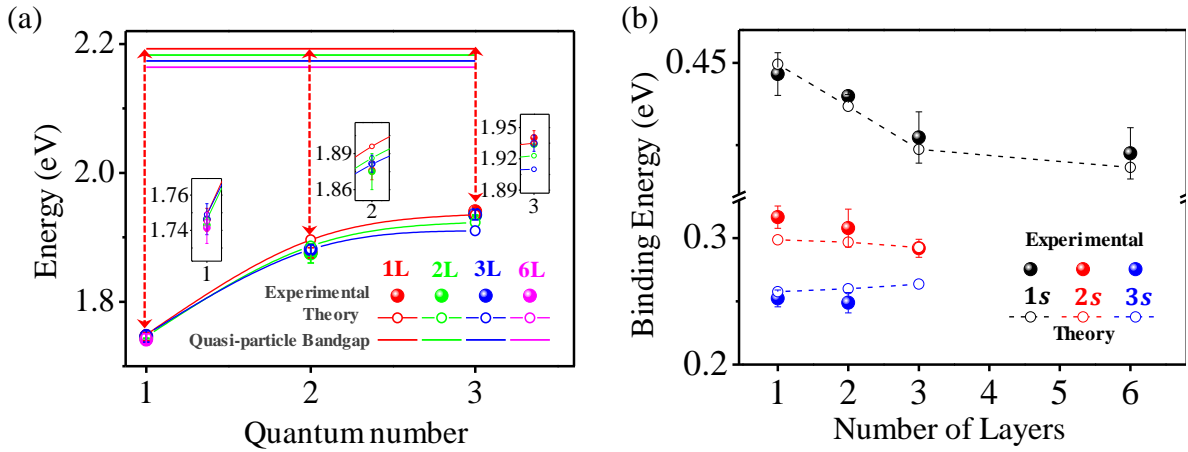

**Figure 6. Thickness dependent binding energy of excitons.** (a) PL emission energy plotted as a function of quantum number. The solid and open symbols represent the emission energies as obtained from photoluminescence experiment ("Experimental") and the Bethe-Salpeter (BS) equation ("Theory"), respectively. The "Theory" values correspond to the bright (quasi-)intra-layer $A_{1s}^{(n)}$ exciton for the n-layer thick film. The continuum, as obtained from BS equation for different layers, is also shown as solid horizontal lines. The binding energy for a given state (1s, 2s, 3s) and sample thickness is extracted by subtracting the emission energy of that state from the corresponding continuum level, as indicated by the dashed red arrows for 1L case. The zoomed-in view of the energy states corresponding to the individual quantum numbers are also shown in the insets. (b) The extracted binding energies from (a), plotted as a function of number of layers. Solid symbols represent experimental values, and open symbols with dashed lines indicate BS equation predicted (Theory) values.

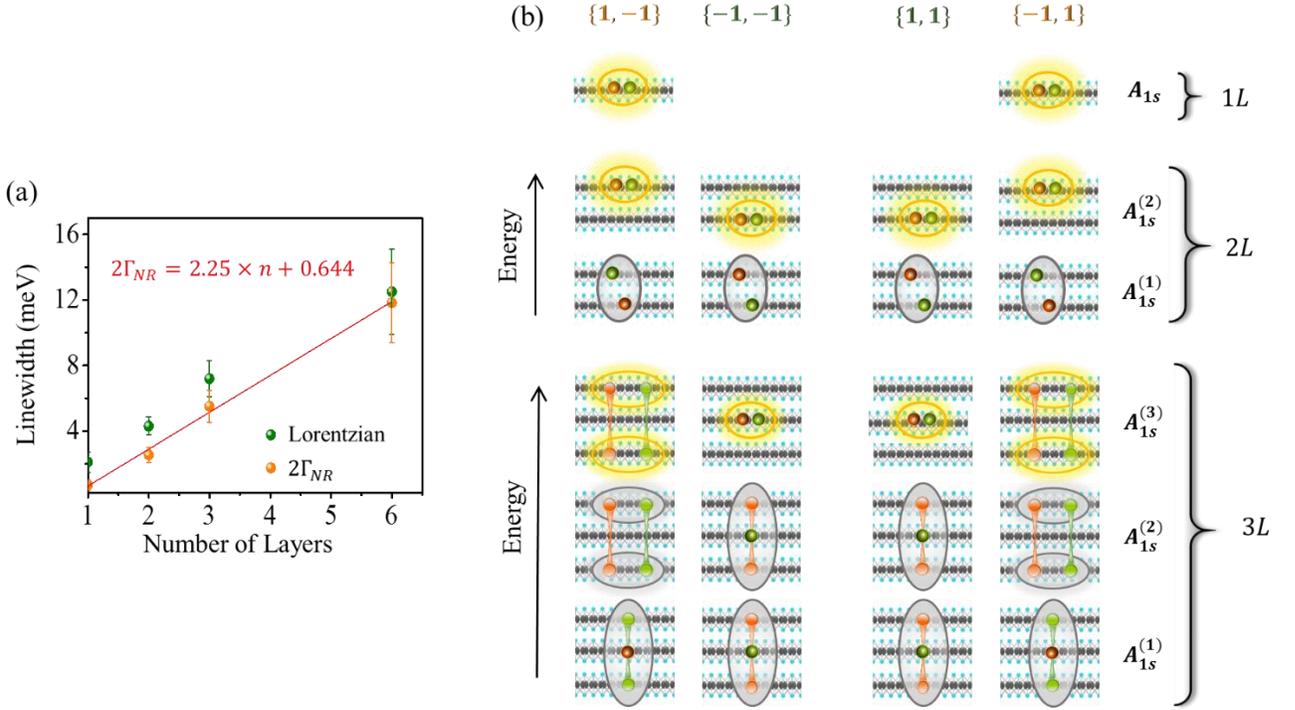

**Figure 7. Exciton scattering to inter-layer dark states.** (a) The experimental Lorentzian linewidth (green symbols) and the corresponding extracted non-radiative broadening $2\Gamma_{NR}$ (orange symbols), as a function of number of layers ($n$). Red line is the fitted function demonstrating linear relationship between non-radiative broadening and $n$. The fitted expression is shown in the inset. (b) Real space layer resolved distribution of the $A_{1s}$ exciton for each spin-valley $\{s_z, \tau_z\}$ configuration for a monolayer (top panel), bilayer (middle panel) and a tri-layer (bottom panel) system.

# Supplemental Material

## Layer degree of freedom for excitons in transition metal dichalcogenides

Sarthak Das[∥], Garima Gupta[∥], and Kausik Majumdar[*]

Department of Electrical Communication Engineering, Indian Institute of Science, Bangalore 560012, India

[∥]These authors contributed equally

[*]Corresponding author, email: kausikm@iisc.ac.in

## S1. k.p quasiparticle Hamiltonian for few-layer TMDCs

The k.p quasi-particle Hamiltonian $H_l$ used in this work for an arbitrary number of layers

$$H_l = \begin{bmatrix} \ddots & \vdots & 0 & \vdots & 0 & 0 & 0 & 0 \\ 0 & \Delta & at_i(\tau_z k_x - ik_y) & 0 & 0 & 0 & 0 & 0 \\ \cdots & at_i(\tau_z k_x + ik_y) & -\tau_z s_z \lambda & 0 & t_\perp & 0 & 0 & 0 \\ 0 & 0 & 0 & \Delta & at_i(\tau_z k_x + ik_y) & 0 & 0 & 0 \\ 0 & t & 0 & at_i(\tau_z k_x - ik_y) & \tau_z s_z \lambda & 0 & t_\perp & 0 \\ 0 & 0 & 0 & 0 & 0 & \Delta & at_i(\tau_z k_x - ik_y) & 0 \\ 0 & 0 & 0 & t & 0 & at_i(\tau_z k_x + ik_y) & -\tau_z s_z \lambda & \cdots \\ 0 & 0 & 0 & 0 & \vdots & 0 & \vdots & \ddots \end{bmatrix} \begin{matrix} \\ \\ \}l-1 \\ \\ \}l \\ \\ \}l+1 \\ \\ \end{matrix}$$

Below, the specific Hamiltonian for bilayer (2L), tri-layer (3L) and four-layer (4L) systems are given.

$$H_{2L} = \begin{bmatrix} \Delta & at_i(\tau_z k_x + ik_y) & 0 & 0 \\ at_i(\tau_z k_x - ik_y) & -\tau_z s_z \lambda & 0 & t_\perp \\ 0 & 0 & \Delta & at_i(\tau_z k_x - ik_y) \\ 0 & t_\perp & at_i(\tau_z k_x + ik_y) & \tau_z s_z \lambda \end{bmatrix}$$

$$H_{3L} = \begin{bmatrix} \Delta & at_i(\tau_z k_x + ik_y) & 0 & 0 & 0 & 0 \\ at_i(\tau_z k_x - ik_y) & -\tau_z s_z \lambda & 0 & t_\perp & 0 & 0 \\ 0 & 0 & \Delta & at_i(\tau_z k_x - ik_y) & 0 & 0 \\ 0 & t_\perp & at_i(\tau_z k_x + ik_y) & \tau_z s_z \lambda & 0 & t_\perp \\ 0 & 0 & 0 & 0 & \Delta & at_i(\tau_z k_x + ik_y) \\ 0 & 0 & 0 & t_\perp & at_i(\tau_z k_x - ik_y) & -\tau_z s_z \lambda \end{bmatrix}$$

$$H_{4L} = \begin{bmatrix} \Delta & at_i(\tau_z k_x + ik_y) & 0 & 0 & 0 & 0 & 0 & 0 \\ at_i(\tau_z k_x - ik_y) & -\tau_z s_z \lambda & 0 & t_\perp & 0 & 0 & 0 & 0 \\ 0 & 0 & \Delta & at_i(\tau_z k_x - ik_y) & 0 & 0 & 0 & 0 \\ 0 & t_\perp & at_i(\tau_z k_x + ik_y) & \tau_z s_z \lambda & 0 & t_\perp & 0 & 0 \\ 0 & 0 & 0 & 0 & \Delta & at_i(\tau_z k_x + ik_y) & 0 & 0 \\ 0 & 0 & 0 & t_\perp & at_i(\tau_z k_x - ik_y) & -\tau_z s_z \lambda & 0 & t_\perp \\ 0 & 0 & 0 & 0 & 0 & 0 & \Delta & at_i(\tau_z k_x - ik_y) \\ 0 & 0 & 0 & 0 & 0 & t_\perp & at_i(\tau_z k_x + ik_y) & \tau_z s_z \lambda \end{bmatrix}$$

Here, $\Delta$ is the monolayer bandgap, $a$ is the lattice constant, $t_i$ is the nearest-neighbour intra-layer hopping, $\lambda$ is the spin-valley coupling for holes in monolayer, $t_\perp$ is the interlayer hopping for holes, $\tau_z$ is the valley degree of freedom ($+1$ for $K$ and $-1$ for $K'$), and $s_z$ is spin degree of freedom ($\pm 1$).

## S2. Decay rate calculation for excitons in an $n$-layer system

The electronic band structure of WSe$_2$ in the basis of $|5d_{z^2}^l\rangle$ and $\frac{1}{\sqrt{2}}\left(|5d_{x^2-y^2}^l\rangle + i\tau|5d_{xy}^l\rangle\right)$ ($l$ denoting the layer index) is obtained by expanding the monolayer $k.p$ Hamiltonian upon incorporating the interlayer coupling of the VBs with the immediate neighbouring layers.

$$\begin{bmatrix} \ddots & \vdots & 0 & \vdots & 0 & 0 & 0 & 0 \\ 0 & \Delta & at_i(\tau_z kx - iky) & 0 & 0 & 0 & 0 & 0 \\ \cdots & at_i(\tau_z kx + iky) & -\tau_z s_z \lambda & 0 & t_\perp & 0 & 0 & 0 \\ 0 & 0 & 0 & \Delta & at_i(\tau_z kx + iky) & 0 & 0 & 0 \\ 0 & t & 0 & at_i(\tau_z kx - iky) & \tau_z s_z \lambda & 0 & t_\perp & 0 \\ 0 & 0 & 0 & 0 & 0 & \Delta & at_i(\tau_z kx - iky) & 0 \\ 0 & 0 & 0 & t & 0 & at_i(\tau_z kx + iky) & -\tau_z s_z \lambda & \cdots \\ 0 & 0 & 0 & 0 & \vdots & 0 & \vdots & \ddots \end{bmatrix} \begin{matrix} \\ \\ \} l-1 \\ \\ \} l \\ \\ \} l+1 \\ \\ \end{matrix}$$

$\Delta$ is the quasi-particle energy gap at $K, K'$, $a$ is the lattice constant and $t$ is the effective hopping integral. $|c, \mathbf{k}\rangle$, $|v, \mathbf{k}\rangle$ denotes the conduction band ($c$) and valence band ($v$) state at $\mathbf{k}$ at energy $\varepsilon_c(\mathbf{k})$ and $\varepsilon_v(\mathbf{k})$.

An exciton state $|X(\mathbf{Q})\rangle$ at a centre of mass momentum $\mathbf{Q}$ is a coherent superposition of electrons and holes from band-pairs $(v, c)$ in an $n$-layer system in the reciprocal space as

$$|X(\mathbf{Q})\rangle = \sum_{v,c,\mathbf{k}} \Upsilon(\mathbf{Q}) |v, \mathbf{k}\rangle |c, \mathbf{k} + \mathbf{Q}\rangle$$

The solution of the Bethe-Salpeter equation for $|X(\mathbf{Q})\rangle$ is the exciton eigen energy $E_s(\mathbf{Q})$ and wavefunction $\Psi_s(\mathbf{Q})$

$$= \sum_{v,c,\mathbf{k}} \lambda^{(s)}_{v,c,\mathbf{Q}}(\mathbf{k}) |v,\mathbf{k}\rangle |c,\mathbf{k}+\mathbf{Q}\rangle,$$

for an exciton in band $s$.

In the dipole approximation, the momentum matrix element given by

$$\mathbf{P}_{v,c,\mathbf{Q}}(\mathbf{k}) = \langle v,\mathbf{k}|\mathbf{p}|c,\mathbf{k}+\mathbf{Q}\rangle = \frac{m}{\hbar}(\varepsilon_c(\mathbf{k}+\mathbf{Q}) - \varepsilon_v(\mathbf{k}))\left\langle v,\mathbf{k}\left|\frac{\partial}{\partial \mathbf{k}}\right|c,\mathbf{k}+\mathbf{Q}\right\rangle$$

Following this, the quantity $\chi_{ex}(\mathbf{Q})$ including contributions from the band-pairs $(v,c)$ and radiative decay rate $\Gamma_R(\mathbf{Q})$ is obtained by

$$\chi_{ex}(\mathbf{Q}) = \sum_{v,c}\left(\int \frac{d^2\mathbf{k}}{(2\pi)^2}\, \mathbf{P}_{v,c,\mathbf{Q}}(\mathbf{k}).\hat{x}\, \lambda^{(s)}_{v,c,\mathbf{Q}}(\mathbf{k})\right)$$

$$\Gamma_R(\mathbf{Q}) = \eta_o \frac{\hbar e^2}{2m_o^2} |\chi_{ex}(\mathbf{Q})|^2 \int_0^\infty dq_z \frac{1}{\sqrt{Q^2+q_z^2}} \times \left(1 + \frac{q_z^2}{Q^2+q_z^2}\right)$$

$$\times \frac{\Gamma(\mathbf{Q})/\pi}{[E_{ex}(\mathbf{Q}) - \hbar c\sqrt{Q^2+q_z^2}] + \Gamma(\mathbf{Q})^2}$$

## S3. Radiative decay rate for excitons in tri-layer (3L) WSe$_2$

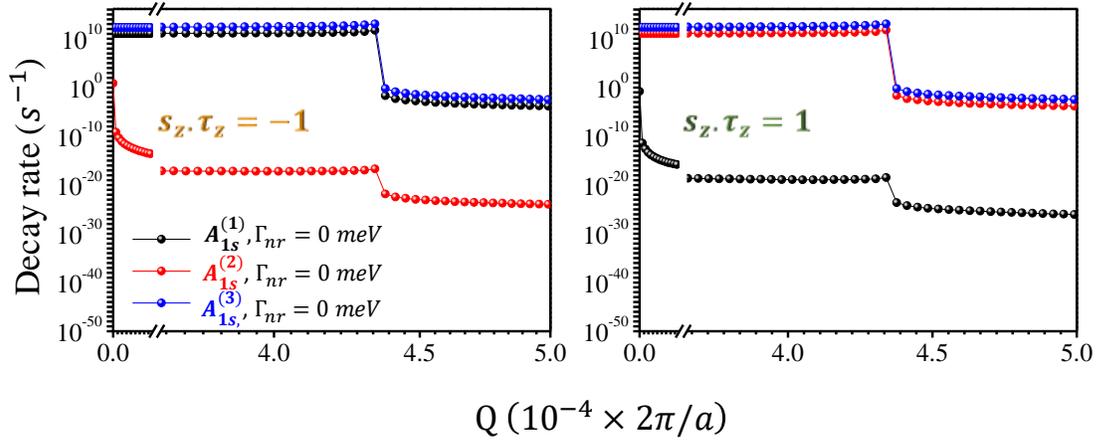

Figure S3. Decay rate calculation for three-layer (3L) excitons with two doubly-degenerate spin valley configurations, taking $\Gamma_{nr} = 0$ meV. While the $A_{1s}^{(3)}$ shows the highest decay rate for both the cases, the other two excitons possess low decay rate.

## S4. Summary of layer dependent excitonic states

Table for experimental excitation energies of excitons*:

| No of layers | Energy of $A_{1s}$ | Energy of $A_{2s}$ | Energy of $A_{3s}$ |
|---|---|---|---|
| 1 | 1.746±0.006 (**1.743**) | 1.876±0.008 (**1.894**) | 1.940±0.006 (**1.935**) |
| 2 | 1.743±0.005 (**1.746**) | 1.875±0.015 (**1.887**) | 1.934±0.007 (**1.922**) |
| 3 | 1.746±0.007 (**1.75**) | 1.881±0.007 (**1.881**) | (**1.91**) |
| 6 | 1.741±0.007 (**1.745**) | - | - |

**\*** all the values are in eV. Values are in parenthesis are calculated values for the bright exciton.

Table for experimentally extracted binding energies of excitons*:

| No of layers | BE of $A_{1s}$ | BE of $A_{2s}$ | BE of $A_{3s}$ |
|---|---|---|---|
| 1 | 0.447±0.006 (**0.45**) | 0.317±0.008 (**0.299**) | 0.252±0.006 (**0.258**) |
| 2 | 0.44±0.005 (**0.437**) | 0.308±0.015 (**0.296**) | 0.249±0.007 (**0.261**) |
| 3 | 0.427±0.007 (**0.423**) | 0.292±0.007 (**0.293**) | (**0.264**) |
| 6 | 0.422±0.007 (**0.418**) | - | - |

**\*** All the values are in eV. Values are in parenthesis are calculated values for the bright exciton.

Summary of calculated energy eigenvalues for excitonic states in bilayer and tri-layer WSe$_2$[*]:

| System | $\tau_z \cdot s_z = +1$ | | | $\tau_z \cdot s_z = -1$ | | |
|---|---|---|---|---|---|---|
| | A$_{1s}$ | A$_{2s}$ | A$_{3s}$ | A$_{1s}$ | A$_{2s}$ | A$_{3s}$ |
| 2L WSe$_2$ | $A_{1s}^{(1)}$ 1.741 | $A_{2s}^{(1)}$ 1.884 | $A_{3s}^{(1)}$ 1.920 | $A_{1s}^{(1)}$ 1.741 | $A_{2s}^{(1)}$ 1.884 | $A_{3s}^{(1)}$ 1.920 |
| | $A_{1s}^{(2)}$ **1.746** | $A_{2s}^{(2)}$ **1.887** | $A_{3s}^{(2)}$ **1.922** | $A_{1s}^{(2)}$ **1.746** | $A_{2s}^{(2)}$ **1.887** | $A_{3s}^{(2)}$ **1.922** |
| 3L WSe$_2$ | $A_{1s}^{(1)}$ 1.746 | $A_{2s}^{(1)}$ 1.877 | $A_{3s}^{(1)}$ 1.908 | $A_{1s}^{(1)}$ 1.746 | $A_{2s}^{(1)}$ 1.877 | $A_{3s}^{(1)}$ 1.908 |
| | $A_{1s}^{(2)}$ 1.746 | $A_{2s}^{(2)}$ 1.877 | $A_{3s}^{(2)}$ 1.908 | $A_{1s}^{(2)}$ 1.75 | $A_{2s}^{(2)}$ 1.881 | $A_{3s}^{(2)}$ 1.909 |
| | $A_{1s}^{(3)}$ **1.75** | $A_{2s}^{(3)}$ **1.881** | $A_{3s}^{(3)}$ **1.909** | $A_{1s}^{(3)}$ **1.75** | $A_{2s}^{(3)}$ **1.881** | $A_{3s}^{(3)}$ **1.909** |

**\*** All the values are in eV. Energy values marked in bold front are used for experimental analysis.

**S5. Layer dependent inhomogeneous broadening of exciton emission linewidth**

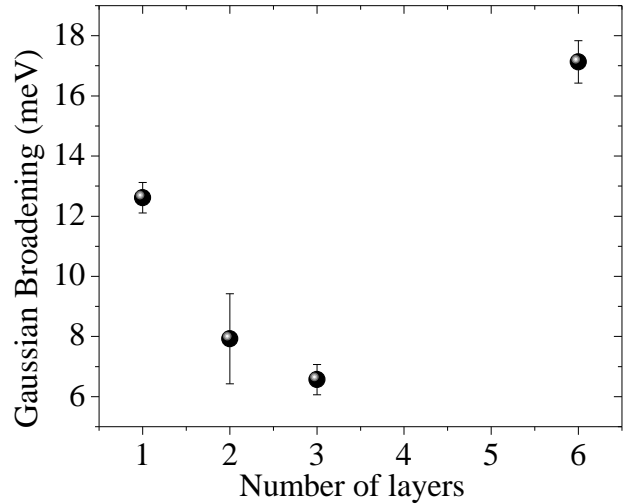

Figure S5. The inhomogeneous (Gaussian) component of the $A_{1s}$ exciton linewidth as a function of number of layers of the WSe$_2$ film, as obtained from the Voigt fitting. The inhomogeneous broadening reduces from 1L to 3L, likely due to improved isolation from SiO$_2$ substrate. However, at 6L, the inhomogeneous broadening is found to be quite large. One possible reason could be due to the layered induced spatial spread of the exciton wavefunction in the different layers, effectively increasing the total inhomogeneity of the sample, with a contribution from each layer.